\documentclass[apjl]{emulateapj}             
\usepackage{graphicx}
\usepackage{txfonts}                        
\usepackage{natbib}
\usepackage{longtable}
\newcommand{\sw}{$Swift$}
\newcommand{\src}{IGR~J17544$-$2619}
\newcommand{\srcigr}{IGR~J17544$-$2619}
\newcommand{\srcxte}{XTE~J1739$-$302}

\def \ATel {The Astronomer's Telegram}

\slugcomment{To appear in ApJ}
\shorttitle{Monitoring SFXTs with $Swift$. III.}
\shortauthors{Sidoli et al.}

\begin{document}
\title{Monitoring Supergiant Fast X-ray Transients with Swift. III. \\
Outbursts of the prototypical SFXTs IGR~J17544$-$2619 and \srcxte. }

\author{L.~Sidoli\altaffilmark{1}, 
P.~Romano\altaffilmark{2}, 
V.~Mangano\altaffilmark{2}, 
G.~Cusumano\altaffilmark{2}, 
S.~Vercellone\altaffilmark{1},  
J.~A.~Kennea\altaffilmark{3},  
A.~Paizis\altaffilmark{1},  
H.~A.~ Krimm\altaffilmark{4,5}, 
D.~N.~Burrows\altaffilmark{3},  
N.~Gehrels\altaffilmark{6} 
}

 \altaffiltext{1}{INAF, Istituto di Astrofisica Spaziale e Fisica Cosmica, 
	Via E.\ Bassini 15,   I-20133 Milano,  Italy}
  \altaffiltext{2}{INAF, Istituto di Astrofisica Spaziale e Fisica Cosmica, 
	Via U.\ La Malfa 153, I-90146 Palermo, Italy} 
   \altaffiltext{3}{Department of Astronomy and Astrophysics, Pennsylvania State 
             University, University Park, PA 16802, USA}
   \altaffiltext{4}{CRESST/Goddard Space Flight Center, Greenbelt, MD, USA}
 \altaffiltext{5}{Universities Space Research Association, Columbia, MD, USA}
   \altaffiltext{6}{NASA/Goddard Space Flight Center, Greenbelt, MD 20771, USA}

\date{\today}

\begin{abstract}
\src\ and \srcxte\ are considered the prototypical sources of the new class
of High Mass X--ray Binaries, the Supergiant Fast X--ray Transients (SFXTs). 
These sources were observed  during bright outbursts
on 2008 March 31 and 2008 April 8, respectively, thanks to an on-going monitoring campaign
we are performing with {\it Swift}, started in October 2007. 
Simultaneous observations with XRT and BAT allowed us to perform
for the first time a broad band spectroscopy of their outbursts.
The X--ray emission is well reproduced with absorbed cutoff powerlaws, 
similar to the typical spectral shape
from accreting pulsars. \src\ shows a significantly harder spectrum 
during the bright flare (where 
a luminosity in excess of 10$^{36}$~erg~s$^{-1}$ is reached)
than during the long-term low level flaring 
activity (10$^{33}$--10$^{34}$~erg~s$^{-1}$), while \srcxte\ displayed the same spectral shape,
within the uncertainties, and 
a higher column density during the flare than in the low level activity.
The light curves of these two SFXTs during the bright flare look similar to those 
observed during recent flares from other  two SFXTs, IGR~J11215--5952 and IGR~J16479--4514,
reinforcing the connection among the members of this class of X--ray sources.
\end{abstract}


\section{Introduction\label{sfxts3:introduction}}

\src\ and \srcxte\ are confirmed members of the new sub-class of High Mass X--ray Binaries,
the Supergiant Fast X--ray Transients (SFXTs), whose members
have been maily discovered with the INTEGRAL satellite (see e.g.\ \citealt{Sguera2005}). 
SFXTs are 
characterized by X--ray transient emission during short (a few hours long) flares
reaching a few 10$^{36}$--10$^{37}$~erg~s$^{-1}$ and they are 
associated with blue supergiant companions 
(e.g.\ \citealt{Negueruela2005a} and  \citealt{Smith2006aa}).
The quiescent state in SFXTs have been observed only in a few sources and is characterized by
a soft spectrum and X--ray luminosity at a level of 10$^{32}$~erg~s$^{-1}$, thus
a very large dynamic range of about 1000--10000 has been observed.
The short duration bright flares are part of a longer accretion phase
at a lower level \citep{Romano2007}. 
When not in outburst, these sources spend most of their
lifetime in accretion at an intermediate (and flaring) level of X--ray luminosity,
of 10$^{33}$--10$^{34}$~erg~s$^{-1}$ \citep{Sidoli2008:sfxts_paperI}.
The spectral properties are reminiscent of those of accreting pulsars,
thus it is likely that several  members of the class are actually hosting neutron
stars, although the spin period has been measured only in two SFXTs 
(AX~J1841.0--0536, \citealt{Bamba2001}; 
IGR~J11215--5952, \citealt{Swank2007:atel999}).

\src\ was discovered \citep{Sunyaev2003} with IBIS/ISGRI on-board INTEGRAL 
on 2003 September 17  during a 2 hour flare reaching 
160~mCrab (18--25~keV). 
During a $Chandra$ observation, both the quiescence level and the onset of
an outburst was caught \citep{zand2005}, observing a dynamic
range as large as 10$^{4}$.
The optical counterpart is an  O9Ib star \citep{Pellizza2006}
located at 3.6 kpc \citep{Rahoui2008}.
Several other bright flares have been observed with INTEGRAL
in 2003, 2004 and 2005 (\citealt{Grebenev2003:17544-2619}, \citealt{Grebenev2004:17544-2619},
\citealt{Sguera2006},  \citealt{Walter2007},  \citealt{Kuulkers2007}),
with flare durations ranging from
0.5 to about 10 hours, reaching peak fluxes of 400~mCrab (20--40 keV).
More recently, two new outbursts were detected with the {\it Swift} satellite,
on 2007 November 8  \citep{Krimm2007:ATel1265}
and on 2008 March 31 \citep{Sidoli2008:atel1454}, 144 days apart. 
The flux at peak observed with {\it Swift}/BAT was 165~mCrab (20--40~keV).
The source was also active on  2007 September 21, with a fainter flaring
emission up to 30--40 mCrab (20--60 keV), as observed with IBIS/ISGRI on-board INTEGRAL
\citep{Kuulkers2007:ATel1266}.


\srcxte\ was discovered with $RXTE$ after a short outburst in August 1997 \citep{Smith1998}, and 
displayed
a spectrum well fitted with a bremsstrahlung model with a temperature of $\sim$22 keV, 
reaching a peak flux of 3.6$\times$10$^{-9}$~erg cm$^{-2}$~s$^{-1}$   (2--25 keV).
Later, several other short flares were observed with $RXTE$/PCA \citep{Smith2006aa}.
The optical counterpart is an  O8I star \citep{Negueruela2006}
located at 2.7 kpc \citep{Rahoui2008}.
Upper limits to the quiescent emission were placed with ASCA observations
\citep{Sakano2002} at a level of $<$1.1$\times$10$^{-12}$~erg cm$^{-2}$~s$^{-1}$.
Bright outbursts (up to 300~mCrab) were detected with IBIS/ISGRI in 2003 March, 
and 2004 March \citep{Sguera2006}.
Frequent flaring activity with INTEGRAL has been reported by \citet{Walter2007}.
Recently, it triggered the {\it Swift} Burst Alert Telescope (BAT).
An immediate slew allowed us to monitor the brightest part of a flare
at soft energies \citep{Romano2008:atel1466}
with the {\it Swift} X-ray Telescope (XRT).
This outburst was also observed by the 
INTEGRAL/JEM-X monitor, which detected a flare starting 5 hours before the flares seen with
{\it Swift} \citep{Chenevez2008}.

\begin{figure}
	\centerline{\includegraphics[width=10cm,height=13cm,angle=0]{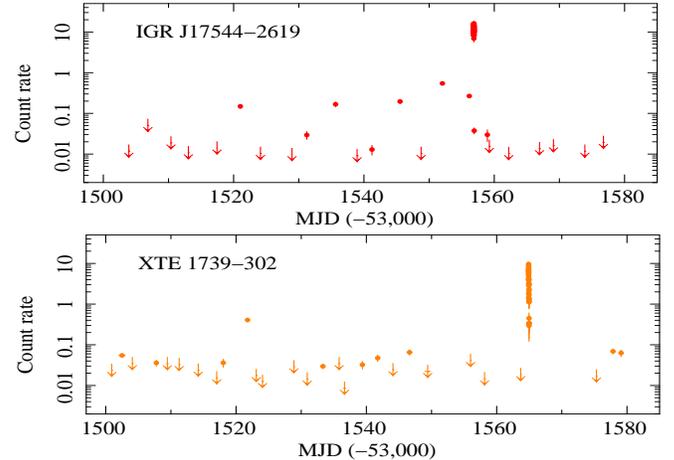}}
	\vspace{-2cm}
                \caption{{\it Swift}/XRT (0.2--10\,keV) light curves of \src\  (upper panel) 
                and of \srcxte\ (lower panel) in 2008, 
		background-subtracted and corrected for pile-up, PSF losses, and vignetting. 
		The data before MJD 54525 from both sources were reported 
                upon in \citet{Sidoli2008:sfxts_paperI}. 
		The downward-pointing arrows are 3-$\sigma$ upper limits. 
               [See the electronic edition of the
                      Journal for a color version of this figure.]
		                }
                \label{sfxts3:fig:lcv_history}
        \end{figure}

Here we report on the detailed analysis of the \sw\ data of two recent outbursts from these
two prototypical SFXTs: the bright flares that 
occurred on 2008 March 31  \citep{Sidoli2008:atel1454} from \src\ 
and  on 2008 April 8  from \srcxte\ \citep{Romano2008:atel1466}.
These observations are part of a monitoring campaign on four SFXTs with \sw, which 
started on 2007 October 26. 
Results on the out-of-outburst emission of the earliest months of \sw/XRT observations are
reported in \citet[][Paper I, see Fig.~\ref{sfxts3:fig:lcv_history}]{Sidoli2008:sfxts_paperI}.
The detailed analysis of the 2008 March 19 outburst of another
SFXT in our monitoring program, IGR~J16479-4514, also caught by {\it Swift},
is reported in \citet[][Paper~II]{Romano2008:sfxts_paperII}.

\section{Observations and data analysis\label{sfxts3:observations}}

\src\ and \srcxte\ triggered the {\it Swift}/BAT on 2008 March 31
20:50:47 UT \citep[image trigger=308224,][]{Sidoli2008:atel1454}, 
and on
2008 April 8
21:28:15 UT \citep[image trigger=308797,][]{Romano2008:atel1466}, 
respectively.
In both occasions, {\it Swift} slewed to the
target, allowing the narrow field instruments (NFIs) to be
pointing at the target $\sim162$ s and $\sim387$\,s after the 
trigger, respectively.
Table~\ref{sfxts3:tab:alldata} reports the log of the {\it Swift} observations
 that were used for this work and which were not listed in \citet{Sidoli2008:sfxts_paperI}.
We note that there were two other outbursts from \src\  seen in
{\it Swift}/BAT (but not reported in literature) on 2007 
September 29 and October 4 (both 8-$\sigma$, 100 mCrab). 
There were smaller flares 
in the BAT both before and after the 2008 April 8 trigger from \srcxte, with some spikes 
reaching $>100$\,mCrab starting at 2008 April 8 16:52 UT and continuing up to 2008 April 9 05:31 UT.

The BAT data were analysed using the standard BAT software  
within FTOOLS ({\tt Heasoft}, v.6.4).
Mask-tagged BAT light curves were created in the standard 4 energy bands, 
15--25, 25--50, 50--100, 100--150 keV, 
 and rebinned to achieve a signal-to-noise ratio (S/N) of 5. 
BAT mask-weighted spectra were extracted over the time 
interval strictly simultaneous with XRT data (see Sect.~3).
Response matrices were generated with {\tt batdrmgen} 
using the latest spectral redistribution matrices. 
For our spectral fitting (XSPEC v11.3.2) 
we  applied an energy-dependent 
systematic error 
vector\footnote{http://heasarc.gsfc.nasa.gov/docs/swift/analysis/bat\_digest.html}.

The XRT data were processed with standard procedures ({\tt xrtpipeline}
v0.11.6), filtering, and screening criteria by using FTOOLS. 
We considered both WT and PC data, 
and selected event grades 0--2 and 0--12, respectively.
When appropriate, we corrected for pile-up. 
To account for the background, we also extracted events within 
source-free regions.
Ancillary response files were generated with {\tt xrtmkarf},
and they account for different extraction regions, vignetting, and
PSF corrections. We used the latest spectral redistribution matrices
(v010) in  the Calibration Database maintained by HEASARC.

Throughout this paper the uncertainties are given at 90\% confidence
level for one interesting parameter unless otherwise stated.
When fitting the broad band spectra during the two bright flares, we 
included factors in the spectral fitting to allow for normalization
uncertainties between the two instruments. 
The constant factors were constrained
to be within their usual ranges during the fitting
(the BAT/XRT constant factor was allowed to vary in the range 0.9--1.1).

\section{Results\label{sfxts3:results}}

\subsection{Light curves}

Figure~\ref{sfxts3:fig:lcv_history}  shows the {\it Swift}/XRT 
0.2--10\,keV light curve of \src\ and \srcxte\ throughout our 2008 monitoring program, 
background-subtracted and corrected for pile-up, PSF losses, 
and vignetting. All data in one segment 
were generally grouped in one point (with the exception of the March 31 and April 8 outbursts). 
The monitoring program started on 2007 October 26 with approximately 
two or three observations per week, with a three-month gap between 2007 November and 2008 
February, when \src\ and \srcxte\ were Sun-constrained. 

\begin{figure}[t]
		\vspace{-1.5truecm}
	\centerline{\includegraphics[width=9.5cm,height=13.5cm,angle=0]{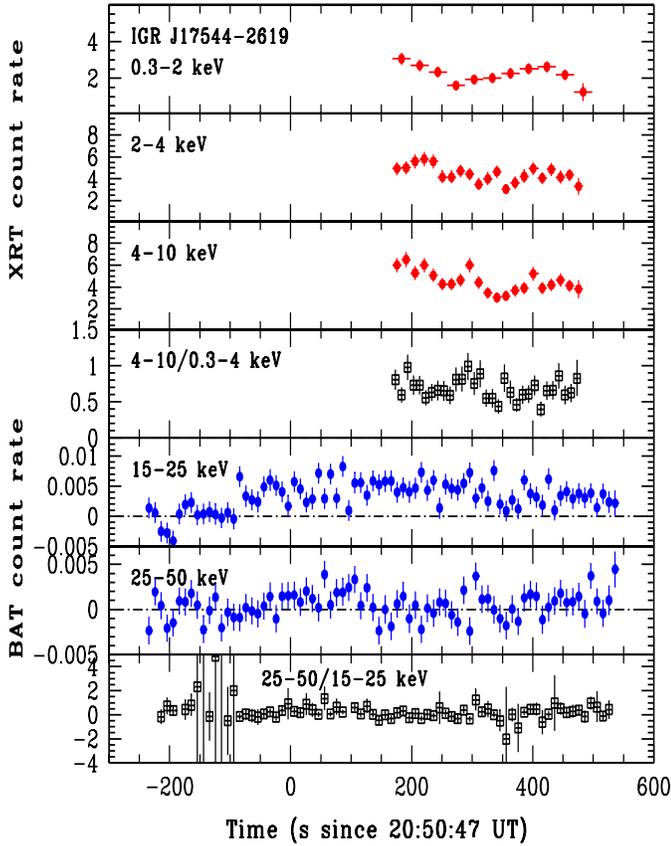}}
		\vspace{-0.7truecm}
                 \caption{XRT and BAT light curves of the 2008 March 31 (MJD 54556) 
                outburst for \src\ 
		in several energy bands in units of count s$^{-1}$ and 
                count s$^{-1}$ detector$^{-1}$, respectively.  
               [See the electronic edition of the
                      Journal for a color version of this figure.]
		}
                \label{fig:lcv_igr}
       \end{figure}

\begin{figure}[t]
	\centerline{\includegraphics[width=9.5cm,height=13.5cm,angle=0]{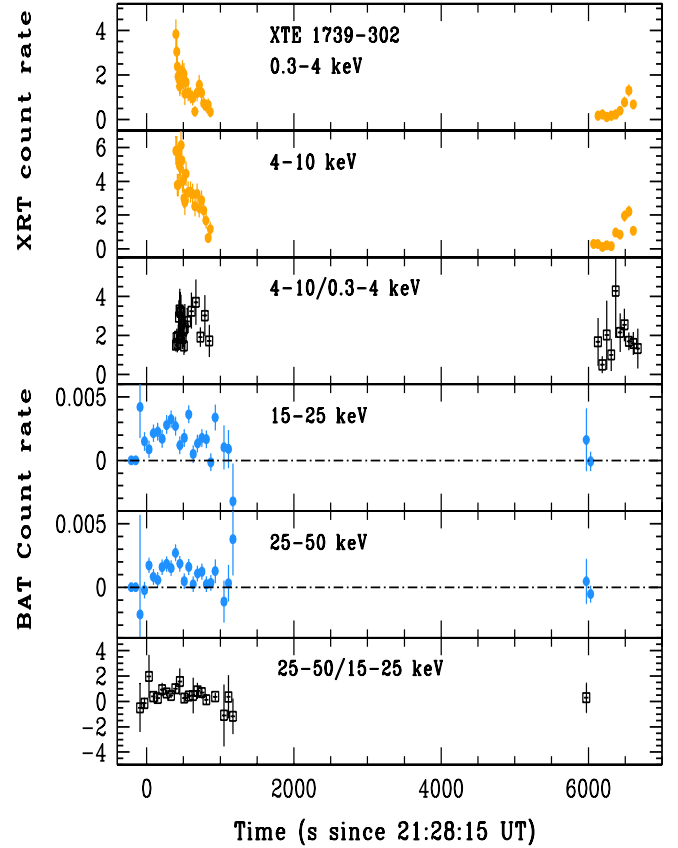}}
		\vspace{-0.7truecm}
                 \caption{XRT and BAT light curves of the 2008 April 8 (MJD 54564) 
                outburst for \srcxte\ 
		in several energy bands in units of count s$^{-1}$ and 
               count s$^{-1}$ detector$^{-1}$, 
		respectively. Note that a second fainter flare has been caught,  about 6000~s after the first one.
               [See the electronic edition of the
                      Journal for a color version of this figure.]
		}
                \label{fig:lcv_xte}
       \end{figure}

Figure~\ref{fig:lcv_igr} and \ref{fig:lcv_xte}  
show the detailed light curves 
in several energy bands during the brightest part of the two outbursts,
together with the 
4--10/0.3--4\,keV, 25--50/15--25\,keV hardness ratios. 
Fitting the \src\ 4--10/0.3--4\,keV hardness ratio as a function of time 
to a constant model yields a value of $0.63\pm0.04$ 
and $\chi^2_{\nu}=1.129$ for 30 degrees of freedom (d.o.f.).
For \srcxte\ we obtain a value of $1.86\pm0.25$ and $\chi^2_{\nu}=0.6$
for 21 degrees of freedom (d.o.f.).

\subsection{Spectroscopy of \srcigr} 

\begin{figure}[th!]
	 \centerline{\includegraphics[width=6.5cm,angle=-90]{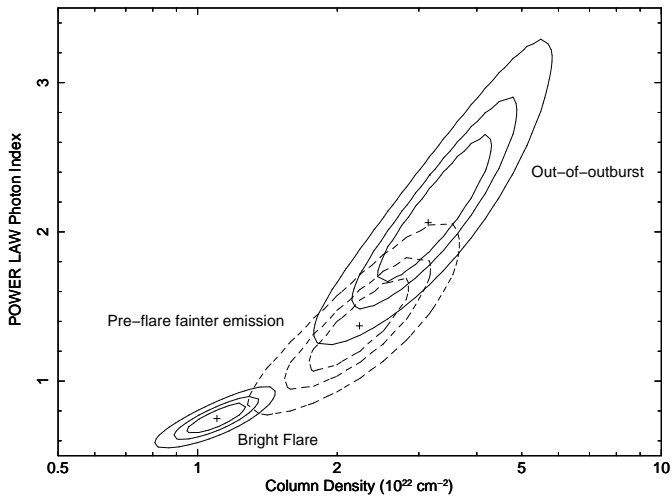}}
                \caption{Spectral parameters 
(absorbed single power law model fit to XRT data only)
derived for \srcigr\ during two {\it Swift}/XRT observations 
(the bright flare on 2008 March 31 (WT data)
and one of the pre-flare observations (dashed contours, PC data, 
obs. 00035056019 in Table~\ref{sfxts3:tab:alldata}), compared to
the average spectrum of the out-of-outburst emission, 
already reported by \citet{Sidoli2008:sfxts_paperI}. 
68\%, 90\% and 99\% confidence level contours are shown.
		                }
                \label{fig:cont}
        \end{figure}

The  XRT/WT spectrum, extracted during the peak of the outburst 
(observation 00308224000, see Table~\ref{sfxts3:tab:alldata}), results in a 
quite hard X--ray emission. 
Adopting an absorbed power law, we obtain a photon index of 
$0.75\pm0.11$, and a high column density, $N_{\rm H}=(1.1\pm0.2)\times 10^{22}$ cm$^{-2}$ 
($\chi^2_{\nu}=0.958$ for 143 d.o.f.). 
The unabsorbed flux in the 2--10\,keV band is $1.2\times10^{-9}$ erg cm$^{-2}$ s$^{-1}$.   
A 3-$\sigma$ upper limit to the equivalent width of an iron line at 
6.7\,keV can be placed at 62\,eV. 
A contour plot is shown in Fig.~\ref{fig:cont} for the single power-law model fit to
the WT spectrum, compared with the out-of-outburst emission \citep{Sidoli2008:sfxts_paperI},
and with one of the observations performed before the flare 
(obs. 00035056019 in Table~\ref{sfxts3:tab:alldata}).
The X--ray spectrum of the fainter emission observed just after the
bright flare (PC data, observation 00035056021) is somewhat softer 
(1.9~$\sigma$, nhp $5.5\times10^{-2}$): indeed, 
fitted using Cash statistics and adopting an absorbed power-law model model, 
the resulting photon index is $1.5_{-0.6}^{+0.7}$ and the absorbing column density 
$N_{\rm H}=(1.0_{-0.6}^{+0.9})\times 10^{22}$ cm$^{-2}$ 
(C-stat$=338.1$ for 63.24\% of 10$^4$ Monte Carlo realizations with statistics 
$<$ C-stat).   
The unabsorbed flux in the 2--10\,keV band is $5\times10^{-12}$ erg cm$^{-2}$ s$^{-1}$. 
A summary of the model parameters can be found in Table~\ref{tab:specs}. 
This table also lists, for comparison, the spectral parameters 
obtained from other  XRT observations reported 
in Table~\ref{sfxts3:tab:alldata}, and performed before the outburst.

\begin{figure}[t]
	 \centerline{\includegraphics[width=9cm,height=12cm,angle=0]{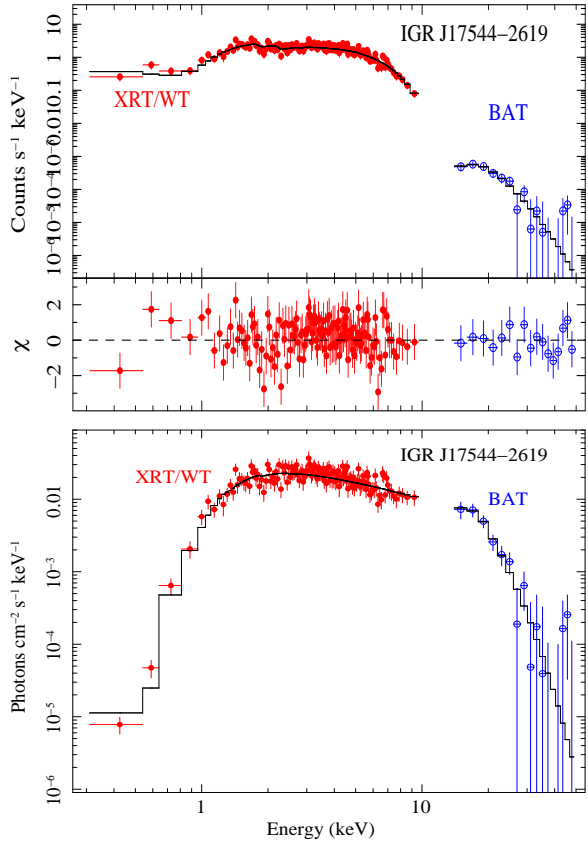}}
                \caption{Spectroscopy of the 2008 March 31 outburst from \src. 
		{\bf Upper panel:} simultaneous BAT and  XRT/WT data 
		fit with an absorbed power law with a high energy cutoff. 
		{\bf Middle panel:} the residuals of the fit (in units of standard deviations). 
                {\bf Lower panel:} the unfolded photon spectra of simultaneous BAT and  XRT/WT data.
		                }
               [See the electronic edition of the
                      Journal for a color version of this figure.]
                \label{sfxts3:fig:meanspec}
        \end{figure}

We fit the simultaneous BAT$+$XRT spectra in the time interval 
168--475\,s since the BAT trigger.
Several models typically used to describe the X--ray emission from 
accreting pulsars in HMXBs were adopted \citep{White1983}. 
For the spectral fitting we considered BAT counts up to 50~keV (above this
energy the statistics is poor).
We report our results in Table~\ref{tab:specigr}, and show an
example of the fits in Fig.~\ref{sfxts3:fig:meanspec}. 
All models result in  a satisfactory deconvolution of the 0.3--50 keV emission,
resulting in a hard powerlaw-like spectrum below 
10\,keV, but a roll over of the high energy emission clearly
emerges when fitting the BAT spectrum together with the XRT 
data.
Very recent theoretical results about the  formation of the spectrum in X--ray pulsars,  
indicate that Comptonization occurs in the shocked gas  in the accretion columns 
onto the neutron star (\citealt{Becker2005} and \citealt{Becker2007}). 
Based on these findings, Comptonization models have  been used in 
describing the spectra observed from several accreting pulsars (see e.g. \citealt{Torrejon2004}, 
\citealt{MasettiOrlandini2006},  \citealt{Ferrigno2008}).
Adopting this more physical description of the spectrum, a Comptonization model ({\sc compTT} in XSPEC,
\citealt{Titarchuk1994}), we obtain a cold  plasma (we assumed 
a spherical geometry for the Comptonization plasma) 
with a well constrained temperature of $\sim$4~keV, and an optical depth of 19$\pm{3}$.

\subsection{Spectroscopy of \srcxte}

\begin{figure}[th!]
	 \centerline{\includegraphics[width=6.5cm,angle=-90]{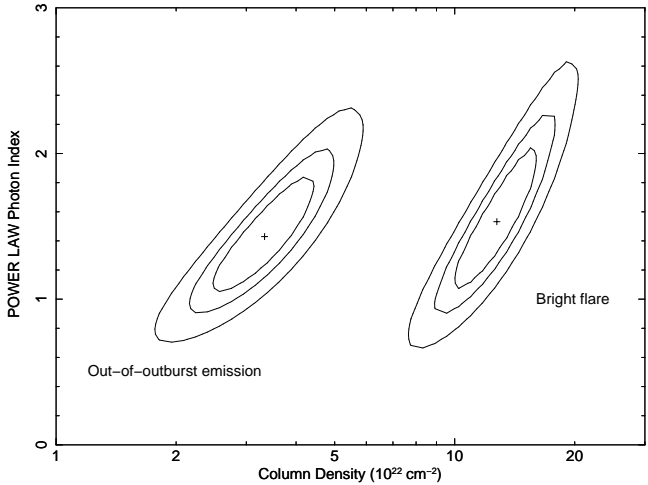}}
                \caption{Comparison of the spectral paramenters (absorbed single power law model)
derived for \srcxte\ during the bright flare on 2008 April 8 (WT data)
and the total spectrum of the out-of-outburst emission  reported in \citet{Sidoli2008:sfxts_paperI}. 
68\%, 90\% and 99\% confidence level contours are shown.
		                }
                \label{fig:contxte}
        \end{figure}

The XRT/WT spectrum (observation 00308797000) extracted during the early part of
the outburst was fit with an absorbed power law, obtaining a photon index of
$1.5_{-0.5}^{+0.6}$, and a high column density, 
$N_{\rm H}$=($13_{-3}^{+4})\times 10^{22}$ cm$^{-2}$
($\chi^2_{\nu}=1.642$ for 35 d.o.f.).
The unabsorbed flux in the 2--10\,keV band is $1.7\times10^{-9}$~erg~cm$^{-2}$~s$^{-1}$. 
A contour plot is shown in Fig.~\ref{fig:contxte} for the single power-law model fit to
the WT spectrum, compared with out-of-outburst emission \citep{Sidoli2008:sfxts_paperI}.
The PC data of the same sequence (observation 00308797000) show a consistent fit: 
photon index of $1.57_{-0.54}^{+0.61}$, $N_{\rm H}$=($13_{-3}^{+4})\times10^{22}$~cm$^{-2}$
($\chi^2_{\nu}=1.082$ for 24 d.o.f.), and an unabsorbed flux in the
2--10\,keV band of
$5\times10^{-10}$ erg cm$^{-2}$ s$^{-1}$.
The model parameters are summarized in Table~\ref{tab:specs}.

\begin{figure}[t]
	 \centerline{\includegraphics[width=9cm,height=12cm,angle=0]{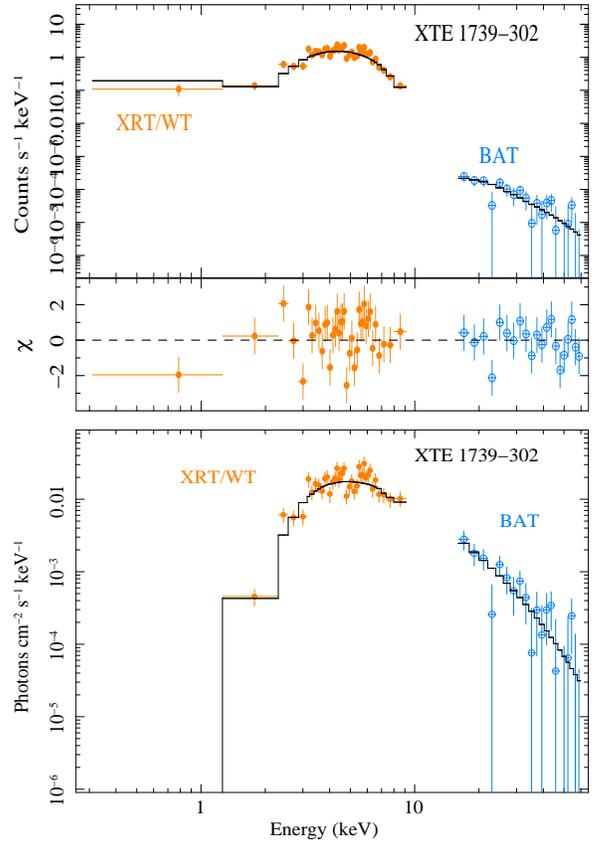}}
                \caption{Spectroscopy of the 2008 April 8 outburst from \srcxte. 
Same as Fig.~\ref{sfxts3:fig:meanspec}.
		                }
                \label{sfxts3:fig:meanspec2}
        \end{figure}

Similarly to the procedure we adopted for \src\, we fit the simultaneous 
XRT$+$BAT spectra of \srcxte\ in the 0.3--10\,keV and the 14--60\,keV energy bands, respectively.
Adopting typical models used to describe the X--ray emission
from HMXBs, as in the case of \src, we obtained the spectral parameters reported 
in Table~\ref{tab:specxte}.  
A steep powerlaw model can reproduce ($\chi^2_{\nu}=1.54$ for 55 d.o.f.) 
the spectrum from soft to hard X-rays,
with a photon index of $2.3 _{-0.1}^{+0.2}$ and an absorbing column density of
$N_{\rm H}$=($18_{-2}^{+3})\times10^{22}$~cm$^{-2}$, although significantly better fits 
are obtained with a cut-off at high energies.
All models (powerlaw with cutoff or Comptonizing plasma model) 
result in equally satisfactory deconvolutions of the 0.3--60 keV emission.
In Fig.~\ref{sfxts3:fig:meanspec2} we show the result obtained adopting a power-law with a 
high energy cut-off.

\section{Discussion\label{sfxts3:discussion}}

Here we report on {\it Swift} observations of \src\ and \srcxte\ during two bright flares,
observed for the first time simultaneously in a broad energy range, from 0.3 to 50--60 keV
(the highest energy where the spectroscopy is meaningful with BAT in these two sources). 
Indeed, before {\it Swift}, the outbursts from these two SFXTs were
mainly observed with INTEGRAL with the high energy detector (E$>$20 keV; e.g. 
\citealt{Sguera2006})
or at softer energy bands 
(\citealt{Smith2006aa}, \citealt{zand2005}).
The only SFXT previously observed simultaneously in a wide X--ray band was IGR~J16479--4514,
during a flare caught with the {\it Swift} satellite \citep{Romano2008:sfxts_paperII}.

The X--ray spectroscopy shows that these two SFXTs, which
are considered the prototypes of this new class of HMXBs, have
different properties during the bright flares.
\src\ is one order of magnitude less absorbed than \srcxte,
and displays a significantly flatter spectrum below 10 keV, with a XRT/WT spectrum
well fitted  with a power-law with a photon index of 0.75$\pm{0.11}$,
compared with the \srcxte\ photon index, which lies in the range 1--2.  
The 1--10 keV spectral properties observed in \src\ during the flare are 
similar to what observed previously with $Chandra$ \citep{zand2005}, where the absorbed
powerlaw fit resulted in a photon index of 0.73$\pm{0.13}$, a column density of
(1.36$\pm{0.22}$)$\times$10$^{22}$~cm$^{-2}$, and a peak flux of $\sim$3$\times$10$^{-9}$~erg~cm$^{-2}$~s$^{-1}$.
 
The broad band analysis shows that \src\ displays a quite sharp cutoff at 18$\pm{2}$~keV
(when using the power law model with a high energy cut-off, {\sc highecut} in  
Table~\ref{tab:specigr}) or a well constrained temperature
for the Comptonizing electrons (in the {\sc comptt} model in XSPEC) 
at 4--5 keV. 
Instead, in \srcxte, a single  power law (photon index of 2.2--2.5)
can  describe the whole spectrum from soft to hard energies.
Part of this difference could be explained by the much higher absorption 
towards the line of sight of \srcxte, which does not allow to constrain
well the low energy part of the power-law model.

        \begin{figure}[th!]
	 \centerline{\includegraphics[width=8.5cm,height=12cm,angle=0]{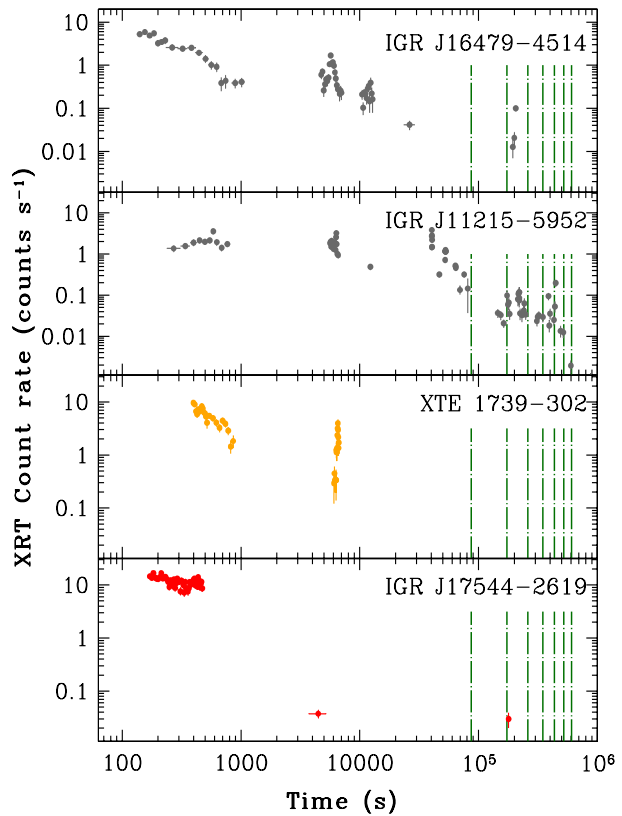}}
                \caption{Light curves of the outbursts of SFXTs followed by {\it Swift}/XRT
referred to their
respective triggers. We show the 2005 outburst of IGR~J16479$-$4514
\citep{Sidoli2008:sfxts_paperI}, which has a better coverage than the one observed in 2008 
\citep{Romano2008:sfxts_paperII}.
The IGR~J11215$-$5952 light curve has an arbitrary start time, since
the source
did not trigger the BAT (the observations were obtained as a ToO; \citet{Romano2007}. 
Note that where no data are plotted, no data were collected. 
For clarity, the time interval between two consecutive 
dashed vertical lines is one day.
		                }
[See the electronic edition of the Journal for a color version of this figure.]
                \label{fig:comp}
        \end{figure}

The observations we are reporting here 
are part of an on-going monitoring campaign of four SFXTs
with {\it Swift}  \citep{Sidoli2008:sfxts_paperI}, which 
started on 2007 October 26.
The two bright flares discussed here  are the first from these two SFXTs,
since the start of the campaign, 
which could be simultaneously covered with both {\it Swift} XRT and BAT.  
The results on the out-of-outburst X--ray emission (below 10 keV) 
have been reported
by \citet{Sidoli2008:sfxts_paperI}, where we find evidence that the accretion is still present,
over long timescales of months, even  outside the bright outbursts. 
Both \srcxte\ and \src\ show evidence that they still accrete matter even outside the outbursts,
at a much fainter (100--1000 times lower) level than during the flares,
with still a large flux variability (at least one order of magnitude).
A complete view of the different luminosity and spectral states of the monitored 
SFXTs will be clearer at the end of the campaign, but it is already possible
to compare the average out-of-outburst emission properties with the spectra during the flares.

Regarding the 0.3--10 keV spectra (fitted with a simple absorbed power-law), 
\srcxte\ appears to be much more absorbed during the flare than during the
out-of-outbust emission (see Fig.~\ref{fig:contxte}), while the
photon index is similar, within the large uncertainties \citep{Sidoli2008:sfxts_paperI}.
Similar changes in the absorbing column density of \srcxte\ have been observed before
with $RXTE$/PCA and ASCA \citep{Smith2006aa}, but during bright outbursts, 
where the $N_{\rm H}$ ranged, from one bright flare to another, 
from 3 to 37$\times10^{22}$~cm$^{-2}$. 
A $Chandra$ observation \citep{Smith2006aa} displaying an unabsorbed 1--10~keV flux 
of $\sim10^{-11}$~erg~cm$^{-2}$~s$^{-1}$,
intermediate between the average out-of-outburst
emission \citep{Sidoli2008:sfxts_paperI} and the bright flare observed here, shows
a hard power law spectrum with a photon index of 0.62$\pm{0.23}$, absorbed
with a column density    $N_{\rm H}$=(4.2$\pm{1.0}$)$\times10^{22}$~cm$^{-2}$, which is compatible with
that of the out-of-outburst emission. 
Thus, in \srcxte, there does not seem to be a clear correlation
between source intensity, spectral hardness and absorbing column density, to date.

Instead, \src\ 
shows a significantly harder spectrum during the flare, and a lower column 
density than during the out-of-outburst phase reported in \citet{Sidoli2008:sfxts_paperI}, obtained summing
together all the XRT data available from 2007 October 27 to 2008 February 28. 
During the out-of-outburst phase, the average observed flux was $\sim$3$\times10^{-12}$~erg~cm$^{-2}$~s$^{-1}$ (2--10 keV),
the powerlaw photon index, $\Gamma$, was 2.1$ ^{+0.6} _{-0.5}$, and the absorbing column density 
N$_{\rm H}$=(3.2 $^{+1.2} _{-0.9}$)$\times10^{22}$~cm$^{-2}$ \citep{Sidoli2008:sfxts_paperI}.
We  also compared the bright flare spectroscopy with the spectrum extracted from one
of the observations obtained a few days before the bright flare 
from \src\ (dashed contours in Fig.~\ref{fig:cont}). 
A hardening of the  \src\ spectrum during the flaring activity is evident.
A similar behaviour was already suggested from the analysis of different 
XMM-Newton observations during a low level flaring activity \citep{Gonzalez2004}, and
from the analysis of the 2004 $Chandra$ observation \citep{zand2005}.
A proper comparison with the INTEGRAL results of a few outbursts from \src\ 
reported by Sguera et al. (2006) cannot be done since the
energy range with INTEGRAL was limited to 20--60~keV. These authors fitted the
20--60~keV spectrum with a thermal bremsstrahlung, which is clearly
not adequate to describe our XRT+BAT spectrum (reduced $\chi^{2}_{\nu}$ of 1.7, for
159 dof).

Different mechanisms have been proposed to explain the bright and 
short duration flaring activity in this new class of sources. 
Some models are related to the structure of the supergiant 
companion wind, involving spherically simmetric clumpy winds 
(see e.g. \citealt{zand2005}, \citealt{Negueruela2008}) 
or  anisotropic winds \citep{Sidoli2007}; other models involve the interaction
of the inflowing wind with the neutron star magnetosphere (see e.g. Bozzo et al. 2008). 
Sidoli et al. (2007) explain the outbursts as being due to enhanced accretion onto
the neutron star when it crosses, moving along the orbit,
an equatorial wind disk component from the supergiant  companion.
Depending on the thickness and truncation of this supposed disk wind 
and on its inclination with respect to the orbital plane of the binary system,
the compact object will cross once or twice in a periodic or quasi-periodic manner
the disk, undergoing outbursts. 
In the framework of this model, the geometry, the structure of this disk wind 
and its inclination with respect to the line
of sight could explain the variability in the 
local absorbing column density, even during different outbursts 
(as observed several times in \srcxte) 
and compared with the low level activity. 
A lower column density during the out-of-outburst activity
could be due to the fact the source is completely outside the denser equatorial wind from
the companion.
In the spherically symmetric clumpy winds model, 
the difference in the observed column
density could be due to the accreting dense clumps. On the other hand, in this case 
the clump matter should remain neutral also in proximity
of the neutron star during bright flares.
We think it is more likely that the absorbing column density is not related with
a neutral accreting matter, but with other clumps or wind structures located probably 
farther away from the compact source.

In Fig.~\ref{fig:comp} we compare the light curves during bright flares from four SFXTs,
all observed with {\it Swift}:
the two reported here from \srcigr\ and  \srcxte, together with the one observed from
IGR~J11215--5952 \citep{Romano2007} and from 
IGR~J16479-4514 \citep{Romano2008:sfxts_paperII}.
All light curves during bright flares look similar, although they 
were observed with a different sampling.
We postpone a more quantitative comparison between the four SFXTs 
(duty cycle, light curve rise time and decay times) 
to a final paper at the end of the on-going observing campaign. 
In any case, it is already evident
that the behaviour of this sample of SFXTs in outburst is similar, 
and that their bright emission 
extends for more than a few hours, contrary to what
originally thought at the time of the discovery of this new class 
of sources \citep[e.g., ][]{Sguera2005}.

The wide band spectra during outbursts display high energy cut-offs (assuming the
model with a power-law modified at high energy by a cutoft, {\sc highecut} in XSPEC), 
although differently constrained in the two sources: in \srcigr\ it is at 18$\pm{2}$~keV,
in \srcxte\ it lies below 13~keV. 
These cut-off ranges are fully consistent with a neutron star magnetic field, B,
of about a 2--3$\times$10$^{12}$~G in the case of \src\ and of 
less than about 2$\times$10$^{12}$~G 
for \srcxte\ \citep{Coburn2002}. These estimates, although not
based on a direct measurement of the magnetic field (which would be possible only
in the case of detection of cyclotron lines), are already difficult
to explain in the framework of the magnetar model recently proposed by \citet{Bozzo2008},
where the magnetic field is at a level of 10$^{14}$~G.
The same is true for other two SFXTs, IGR~J11215--5952 \citep{Sidoli2007} 
and IGR~J16479--4514 \citep{Romano2008:sfxts_paperII}.

{\it Facilities:} \facility{{\it Swift}}.

\acknowledgements
We thank the \sw\ team for making these observations possible,
in particular Scott Barthelmy (for his invaluable help with BAT),
the duty scientists,  and science planners.
PR thanks INAF-IASF Milano, where part of the work was carried out, for their kind hospitality. 
HK was supported by the \sw\ project. 
This work was supported in Italy by MIUR grant 2005-025417 and contract ASI/INAF I/023/05/0, 
I/008/07/0 and I/088/06/0, and at PSU by NASA contract NAS5-00136.


\begin{thebibliography}{35}
\expandafter\ifx\csname natexlab\endcsname\relax\def\natexlab#1{#1}\fi

\bibitem[{{Bamba} {et~al.}(2001){Bamba}, {Yokogawa}, {Ueno}, {Koyama}, \&
  {Yamauchi}}]{Bamba2001}
{Bamba}, A., {Yokogawa}, J., {Ueno}, M., {Koyama}, K., \& {Yamauchi}, S. 2001,
  \pasj, 53, 1179

\bibitem[{{Becker} \& {Wolff}(2005)}]{Becker2005}
{Becker}, P.~A., \& {Wolff}, M.~T. 2005, \apj, 630, 465

\bibitem[{{Becker} \& {Wolff}(2007)}]{Becker2007}
---. 2007, \apj, 654, 435

\bibitem[{{Bozzo} {et~al.}(2008){Bozzo}, {Falanga}, \& {Stella}}]{Bozzo2008}
{Bozzo}, E., {Falanga}, M., \& {Stella}, L. 2008, ArXiv e-prints, 805, \apj, in press

\bibitem[{{Chenevez} {et~al.}(2008){Chenevez}, {Beckmann}, {Kuulkers}, {Bird},
  {Brandt}, {Courvoisier}, {Domingo}, {Ebisawa}, {Jonker}, {Kretschmar},
  {Markwardt}, {Oosterbroek}, {Paizis}, {Risquez}, {Sanchez-Fernandez}, {Shaw},
  \& {Wijnands}}]{Chenevez2008}
{Chenevez}, J. {et~al.} 2008, The Astronomer's Telegram, 1471, 1

\bibitem[{{Coburn} {et~al.}(2002){Coburn}, {Heindl}, {Rothschild}, {Gruber},
  {Kreykenbohm}, {Wilms}, {Kretschmar}, \& {Staubert}}]{Coburn2002}
{Coburn}, W., {Heindl}, W.~A., {Rothschild}, R.~E., {Gruber}, D.~E.,
  {Kreykenbohm}, I., {Wilms}, J., {Kretschmar}, P., \& {Staubert}, R. 2002,
  \apj, 580, 394

\bibitem[{{Ferrigno} {et~al.}(2008){Ferrigno}, {Segreto}, {Mineo},
  {Santangelo}, \& {Staubert}}]{Ferrigno2008}
{Ferrigno}, C., {Segreto}, A., {Mineo}, T., {Santangelo}, A., \& {Staubert}, R.
  2008, \aap, 479, 533

\bibitem[{{Gonz{\'a}lez-Riestra} {et~al.}(2004){Gonz{\'a}lez-Riestra},
  {Oosterbroek}, {Kuulkers}, {Orr}, \& {Parmar}}]{Gonzalez2004}
{Gonz{\'a}lez-Riestra}, R., {Oosterbroek}, T., {Kuulkers}, E., {Orr}, A., \&
  {Parmar}, A.~N. 2004, \aap, 420, 589

\bibitem[{{Grebenev} {et~al.}(2003){Grebenev}, {Lutovinov}, \&
  {Sunyaev}}]{Grebenev2003:17544-2619}
{Grebenev}, S.~A., {Lutovinov}, A.~A., \& {Sunyaev}, R.~A. 2003, \ATel, 192

\bibitem[{{Grebenev} {et~al.}(2004){Grebenev}, {Rodriguez}, {Westergaard},
  {Sunyaev}, \& {Oosterbroek}}]{Grebenev2004:17544-2619}
{Grebenev}, S.~A., {Rodriguez}, J., {Westergaard}, N.~J., {Sunyaev}, R.~A., \&
  {Oosterbroek}, T. 2004, \ATel, 252

\bibitem[{{in't Zand}(2005)}]{zand2005}
{in't Zand}, J.~J.~M. 2005, \aap, 441, L1

\bibitem[{{Krimm} {et~al.}(2007){Krimm}, {Barthelmy}, {Barbier}, {Cummings},
  {Fenimore}, {Gehrels}, {Markwardt}, {Palmer}, {Parsons}, {Sakamoto}, {Sato},
  {Stamatikos}, \& {Tueller}}]{Krimm2007:ATel1265}
{Krimm}, H.~A. {et~al.} 2007, \ATel, 1265

\bibitem[{{Kuulkers} {et~al.}(2007{\natexlab{a}}){Kuulkers}, {Oneca}, {Brandt},
  {Shaw}, {Beckmann}, {Chenevez}, {Courvoisier}, {Domingo}, {Ebisawa},
  {Jonker}, {Kretschmar}, {Markwardt}, {Oosterbroek}, {Paizis},
  {Sanchez-Fernandez}, \& {Wijnands}}]{Kuulkers2007:ATel1266}
{Kuulkers}, E. {et~al.} 2007{\natexlab{a}}, The Astronomer's Telegram, 1266

\bibitem[{{Kuulkers} {et~al.}(2007{\natexlab{b}}){Kuulkers}, {Shaw}, {Paizis},
  {Chenevez}, {Brandt}, {Courvoisier}, {Domingo}, {Ebisawa}, {Kretschmar},
  {Markwardt}, {Mowlavi}, {Oosterbroek}, {Orr}, {R{\'{\i}}squez},
  {Sanchez-Fernandez}, \& {Wijnands}}]{Kuulkers2007}
---. 2007{\natexlab{b}}, \aap, 466, 595

\bibitem[{{Masetti} {et~al.}(2006){Masetti}, {Orlandini}, {dal Fiume}, {del
  Sordo}, {Amati}, {Frontera}, {Palazzi}, \&
  {Santangelo}}]{MasettiOrlandini2006}
{Masetti}, N., {Orlandini}, M., {dal Fiume}, D., {del Sordo}, S., {Amati}, L.,
  {Frontera}, F., {Palazzi}, E., \& {Santangelo}, A. 2006, \aap, 445, 653

\bibitem[{{Negueruela} {et~al.}(2006{\natexlab{a}}){Negueruela}, {Smith},
  {Harrison}, \& {Torrej{\'o}n}}]{Negueruela2006}
{Negueruela}, I., {Smith}, D.~M., {Harrison}, T.~E., \& {Torrej{\'o}n}, J.~M.
  2006{\natexlab{a}}, \apj, 638, 982

\bibitem[{{Negueruela} {et~al.}(2006{\natexlab{b}}){Negueruela}, {Smith},
  {Reig}, {Chaty}, \& {Torrej{\'o}n}}]{Negueruela2005a}
{Negueruela}, I., {Smith}, D.~M., {Reig}, P., {Chaty}, S., \& {Torrej{\'o}n},
  J.~M. 2006{\natexlab{b}}, in Proceedings of the ``The X-ray Universe 2005'',
  26-30 September 2005, El Escorial, Madrid, Spain. Ed. by A. Wilson. ESA
  SP-604, Volume 1, Noordwijk: ESA Publications Division, ISBN 92-9092-915-4,
  2006, ed. A.~{Wilson}, 165

\bibitem[{{Negueruela} {et~al.}(2008){Negueruela}, {Torrejon}, {Reig}, {Ribo},
  \& {Smith}}]{Negueruela2008}
{Negueruela}, I., {Torrejon}, J.~M., {Reig}, P., {Ribo}, M., \& {Smith}, D.~M.
  2008, ArXiv e-prints, 801

\bibitem[{{Pellizza} {et~al.}(2006){Pellizza}, {Chaty}, \&
  {Negueruela}}]{Pellizza2006}
{Pellizza}, L.~J., {Chaty}, S., \& {Negueruela}, I. 2006, \aap, 455, 653

\bibitem[{{Rahoui} {et~al.}(2008){Rahoui}, {Chaty}, {Lagage}, \&
  {Pantin}}]{Rahoui2008}
{Rahoui}, F., {Chaty}, S., {Lagage}, P.-O., \& {Pantin}, E. 2008, ArXiv
  e-prints 0802.1770

\bibitem[{{Romano} {et~al.}(2008{\natexlab{a}}){Romano}, {Sidoli}, {Mangano},
  {Krimm}, {Barthelmy}, , {Burrows}, {Cusumano}, {Gehrels}, {Kennea}, {Paizis},
  {Tueller}, \& {Vercellone}}]{Romano2008:atel1466}
{Romano}, P. {et~al.} 2008{\natexlab{a}}, \ATel, 1466

\bibitem[{{Romano} {et~al.}(2007){Romano}, {Sidoli}, {Mangano}, {Mereghetti},
  \& {Cusumano}}]{Romano2007}
{Romano}, P., {Sidoli}, L., {Mangano}, V., {Mereghetti}, S., \& {Cusumano}, G.
  2007, \aap, 469, L5

\bibitem[{{Romano} {et~al.}(2008{\natexlab{b}}){Romano}, {Sidoli}, {Mangano},
  {Vercellone}, {Kennea}, {Cusumano}, {Krimm}, {Burrows}, \&
  {Gehrels}}]{Romano2008:sfxts_paperII}
{Romano}, P. {et~al.} 2008{\natexlab{b}}, \apjl, 680, L137 (Paper~II)

\bibitem[{{Sakano} {et~al.}(2002){Sakano}, {Koyama}, {Murakami}, {Maeda}, \&
  {Yamauchi}}]{Sakano2002}
{Sakano}, M., {Koyama}, K., {Murakami}, H., {Maeda}, Y., \& {Yamauchi}, S.
  2002, \apjs, 138, 19

\bibitem[{{Sguera} {et~al.}(2005){Sguera}, {Barlow}, {Bird}, {Clark}, {Dean},
  {Hill}, {Moran}, {Shaw}, {Willis}, {Bazzano}, {Ubertini}, \&
  {Malizia}}]{Sguera2005}
{Sguera}, V. {et~al.} 2005, \aap, 444, 221

\bibitem[{{Sguera} {et~al.}(2006){Sguera}, {Bazzano}, {Bird}, {Dean},
  {Ubertini}, {Barlow}, {Bassani}, {Clark}, {Hill}, {Malizia}, {Molina}, \&
  {Stephen}}]{Sguera2006}
---. 2006, \apj, 646, 452

\bibitem[{{Sidoli} {et~al.}(2008{\natexlab{a}}){Sidoli}, {Romano}, {Mangano},
  {Kennea}, {Cusumano}, {Campana}, {Vercellone}, {Burrows}, \&
  {Gehrels}}]{Sidoli2008:atel1454}
{Sidoli}, L. {et~al.} 2008{\natexlab{a}}, \ATel, 1454

\bibitem[{{Sidoli} {et~al.}(2008{\natexlab{b}}){Sidoli}, {Romano}, {Mangano},
  {Pellizzoni}, {Kennea}, {Cusumano}, {Vercellone}, {Paizis}, {Burrows}, \&
  {Gehrels}}]{Sidoli2008:sfxts_paperI}
---. 2008{\natexlab{b}}, \apj, in press, astro-ph/0805.1808 (Paper~I)

\bibitem[{{Sidoli} {et~al.}(2007){Sidoli}, {Romano}, {Mereghetti}, {Paizis},
  {Vercellone}, {Mangano}, \& {G{\"o}tz}}]{Sidoli2007}
{Sidoli}, L., {Romano}, P., {Mereghetti}, S., {Paizis}, A., {Vercellone}, S.,
  {Mangano}, V., \& {G{\"o}tz}, D. 2007, \aap, 476, 1307

\bibitem[{{Smith} {et~al.}(2006){Smith}, {Heindl}, {Markwardt}, {Swank},
  {Negueruela}, {Harrison}, \& {Huss}}]{Smith2006aa}
{Smith}, D.~M., {Heindl}, W.~A., {Markwardt}, C.~B., {Swank}, J.~H.,
  {Negueruela}, I., {Harrison}, T.~E., \& {Huss}, L. 2006, \apj, 638, 974

\bibitem[{{Smith} {et~al.}(1998){Smith}, {Main}, {Marshall}, {Swank}, {Heindl},
  {Leventhal}, {in 't Zand}, \& {Heise}}]{Smith1998}
{Smith}, D.~M., {Main}, D., {Marshall}, F., {Swank}, J., {Heindl}, W.~A.,
  {Leventhal}, M., {in 't Zand}, J.~J.~M., \& {Heise}, J. 1998, \apjl, 501,
  L181+

\bibitem[{{Sunyaev} {et~al.}(2003){Sunyaev}, {Grebenev}, {Lutovinov},
  {Rodriguez}, {Mereghetti}, {Gotz}, \& {Courvoisier}}]{Sunyaev2003}
{Sunyaev}, R.~A., {Grebenev}, S.~A., {Lutovinov}, A.~A., {Rodriguez}, J.,
  {Mereghetti}, S., {Gotz}, D., \& {Courvoisier}, T. 2003, The Astronomer's
  Telegram, 190, 1

\bibitem[{{Swank} {et~al.}(2007){Swank}, {Smith}, \&
  {Markwardt}}]{Swank2007:atel999}
{Swank}, J., {Smith}, D., \& {Markwardt}, C. 2007, \ATel, 997

\bibitem[{{Titarchuk}(1994)}]{Titarchuk1994}
{Titarchuk}, L. 1994, \apj, 434, 570

\bibitem[{{Torrej{\'o}n} {et~al.}(2004){Torrej{\'o}n}, {Kreykenbohm}, {Orr},
  {Titarchuk}, \& {Negueruela}}]{Torrejon2004}
{Torrej{\'o}n}, J.~M., {Kreykenbohm}, I., {Orr}, A., {Titarchuk}, L., \&
  {Negueruela}, I. 2004, \aap, 423, 301

\bibitem[{{Walter} \& {Zurita Heras}(2007)}]{Walter2007}
{Walter}, R., \& {Zurita Heras}, J. 2007, \aap, 476, 335

\bibitem[{{White} {et~al.}(1983){White}, {Swank}, \& {Holt}}]{White1983}
{White}, N.~E., {Swank}, J.~H., \& {Holt}, S.~S. 1983, \apj, 270, 711

\end{thebibliography}


\begin{deluxetable}{lllllr}
  \tabletypesize{\scriptsize}
  \tablewidth{0pc} 	      	
  \tablecaption{Observation log.\label{sfxts3:tab:alldata}} 
  \tablehead{
\colhead{Name} & \colhead{Sequence\tablenotemark{a}} & \colhead{Instrument/Mode} & \colhead{Start time (UT)} &  \colhead{End time (UT)} &  
               \colhead{Net Exposure\tablenotemark{b}} \\
\colhead{}       & \colhead{}            & \colhead{}    & \colhead{(yyyy-mm-dd hh:mm:ss)} & \colhead{(yyyy-mm-dd hh:mm:ss)} &  \colhead{(s)} \\
\colhead{(1)}    & \colhead{(2)}         & \colhead{(3)} & \colhead{(4)}         & \colhead{(5)} & \colhead{(6)}
}
  \startdata
IGR~J17544$-$2619 & & & & & \\ 
		& 00035056012	  &XRT/PC   &2008-03-03 22:43:07       &2008-03-03 23:00:57	&	1071	\\
		& 00035056013	  &XRT/PC   &2008-03-06 03:52:22       &2008-03-06 07:13:58	&	1180	\\
		& 00035056014	  &XRT/PC   &2008-03-10 13:54:43       &2008-03-10 17:13:58	&	627	\\
		& 00035056015	  &XRT/PC   &2008-03-13 22:06:37       &2008-03-13 23:42:56	&	1296	\\
		& 00035056016	  &XRT/PC   &2008-03-16 04:43:43       &2008-03-16 06:33:56	&	1701	\\
		& 00035056017	  &XRT/PC   &2008-03-20 12:57:18       &2008-03-20 13:07:58	&	639	\\
		& 00035056018	  &XRT/PC   &2008-03-23 18:20:14       &2008-03-23 18:36:56	&	1002	\\
		& 00035056019	  &XRT/PC   &2008-03-27 00:45:04       &2008-03-27 01:02:57	&	1073	\\
		& 00035056020	  &XRT/PC   &2008-03-31 03:07:59       &2008-03-31 04:52:57	&	898	\\
		& 00308224000	  & BAT/evt &2008-03-31 20:46:48       &2008-03-31 21:06:50	&       1202   \\
		& 00308224000	  &XRT/WT   &2008-03-31 20:53:35       &2008-03-31 20:58:57	&	306	\\
		& 00035056021	  &XRT/PC   &2008-03-31 21:52:49       &2008-03-31 22:17:41	&	1492	\\
		& 00035056022	  &XRT/PC   &2008-04-02 22:29:22       &2008-04-02 22:42:57	&	815	\\
		& 00035056024	  &XRT/PC   &2008-04-03 05:05:00       &2008-04-03 06:47:56	&	1098	\\
		& 00035056025	  &XRT/PC   &2008-04-06 05:12:15       &2008-04-06 05:28:58     &       1000    \\
		& 00035056026     &XRT/PC   &2008-04-10 21:46:21     & 2008-04-10 23:30:56     &       1027    \\
		& 00035056027     &XRT/PC   &2008-04-13 01:12:20     & 2008-04-13 02:54:28     &       765    \\
		& 00035056028	  &XRT/PC   &2008-04-17 20:45:21     & 2008-04-17 22:34:57     &       1030    \\
		& 00035056029	  &XRT/PC   &2008-04-20 16:21:01     & 2008-04-20 19:36:56     &       888     \\
XTE~J1739$-$302   & & & & & \\ 
	%
	&00030987016	&XRT/PC    &	    2008-03-03 21:06:00     &	    2008-03-03 21:22:58     &	    1016    \\
	&00030987017	&XRT/PC    &	    2008-03-05 21:21:59     &	    2008-03-05 22:55:57     &	    1244    \\
	&00030987018	&XRT/PC    &	    2008-03-08 04:03:06     &	    2008-03-08 12:12:57     &	    2821    \\
	&00030987019	&XRT/PC    &	    2008-03-10 18:43:52     &	    2008-03-10 20:25:56     &	    647     \\
	&00030987020	&XRT/PC    &	    2008-03-11 12:16:41     &	    2008-03-11 18:54:58     &	    2908    \\
	&00030987021	&XRT/PC    &	    2008-03-14 09:21:55     &	    2008-03-14 09:40:50     &	    1108    \\
	&00030987022	&XRT/PC    &	    2008-03-16 17:34:59     &	    2008-03-16 17:51:57     &	    1018    \\
	&00030987023	&XRT/PC    &	    2008-03-19 00:28:15     &	    2008-03-19 03:45:50     &	    672     \\
	&00030987024	&XRT/PC    &	    2008-03-21 05:18:20     &	    2008-03-22 00:41:56     &	    1043    \\
	&00030987025	&XRT/PC    &	    2008-03-23 23:14:57     &	    2008-03-24 20:17:57     &	    1257    \\
	&00030987026	&XRT/PC    &	    2008-03-30 23:41:20     &	    2008-03-30 23:57:58     &	    996     \\
	&00030987027	&XRT/PC    &	    2008-04-02 03:08:26     &	    2008-04-02 03:24:56     &	    990     \\
	&00030987028	&XRT/PC    &	    2008-04-07 16:27:21     &	    2008-04-07 16:43:56     &	    995     \\
        &00308797000	&BAT/evt   &        2008-04-08 21:24:16     &       2008-04-08 23:09:11     &      1735       \\
	&00308797000	&XRT/WT    &	    2008-04-08 21:34:47     &	    2008-04-08 23:09:19     &	    128     \\
	&00308797000	&XRT/PC    &	    2008-04-08 21:36:54     &	    2008-04-08 23:19:19     &	    938     \\
	&00030987029	&XRT/PC    &	    2008-04-19 06:26:23     &	    2008-04-19 08:09:56     &	    908     \\
	&00030987030	&XRT/PC    &	    2008-04-21 19:28:21     &	    2008-04-21 21:15:57     &	    1146    \\
	&00030987031    &XRT/PC    &        2008-04-23 02:03:03     &       2008-04-23 02:14:58     &       689   	    
\enddata 
  \tablenotetext{a}{The previous observations are listed in \citet{Sidoli2008:sfxts_paperI}.}
  \tablenotetext{b}{The exposure time is spread over several snapshots  
	(single continuous pointings at the target) during each observation.}
  \end{deluxetable}


\begin{deluxetable}{lllrrrr}
  \tabletypesize{\scriptsize}
  \tablewidth{0pc} 	      	
  \tablecaption{Absorbed power-law spectral fits of XRT data. \label{tab:specs}}
  \tablehead{
  \colhead{Name}& \colhead{Date}         & \colhead{Spectrum} &  \colhead{$N_{\rm H}$}           &  \colhead{$\Gamma$} &  \colhead{$\chi^{2}_{\nu}$ (dof)} & \colhead{Obs. Flux (1--10 keV)} \\
  \colhead{}    & \colhead{(yyyy-mm-dd)} & \colhead{}         &  \colhead{($10^{22}$ cm$^{-2}$)} &  \colhead{}         &  \colhead{}  & \colhead{($10^{-10}$ erg cm$^{-2}$ s$^{-1}$)} 
}
  \startdata
  \srcigr 	&2008-03-10 & XRT/PC	&$2.0_{-0.8}^{+1.0}$    &$2.2_{-0.7}^{+0.8}$ &334.7 (71.5\,\%) &  $\sim$0.1 \\ 
		&2008-03-20 & XRT/PC	&$2.2_{-0.9}^{+1.1}$    &$2.0_{-0.6}^{+0.7}$ &388.8 (77.2\,\%) &  $\sim$0.1 \\ 
		&2008-03-27 & XRT/PC	&$2.2_{-0.6}^{+0.7}$      &$1.4_{-0.3}^{+0.4}$ &0.854 (19)	      &  0.5 \\ 
		&2008-03-31 & XRT/PC	&$3.6_{-1.0}^{+1.3}$      &$2.3_{-0.5}^{+0.6}$ &551.0 (82.3\,\%) &  $\sim$0.2 \\ 
                &2008-03-31 & XRT/WT	&$1.1\pm0.2$		&$0.75\pm0.11$		& 0.958 (143)                        & 11 \\ 
         	&2008-03-31 & XRT/PC	&$1.0_{-0.6}^{+0.9}$	&$1.5_{-0.6}^{+0.7}$	& 338.1 (63.24\,\%)\tablenotemark{a} & $\sim$0.04 \\ 

  \srcxte    	&2008-04-08 & XRT/WT	&$13_{-3}^{+4}$		&$1.5_{-0.5}^{+0.6}$	& 1.642 (35)  & 9 \\
   		&2008-04-08 & XRT/PC	&$12_{-3}^{+4}$		&$1.6_{-0.5}^{+0.6}$	& 1.082 (24)  & 3

  \smallskip
  \enddata 
  \tablenotetext{a}{Cash statistics and percentage of Monte Carlo realizations with statistic $<$C-stat. }
  \end{deluxetable}

\begin{deluxetable}{lrrrrrrr}
  \tabletypesize{\scriptsize}
  \tablewidth{0pc} 	      	
  \tablecaption{Spectral fits of simultaneous XRT and BAT data of \srcigr.\label{tab:specigr}}
  \tablehead{
\colhead{\srcigr} & \colhead{} & \colhead{} & \colhead{} &  \colhead{} &  \colhead{} &  \colhead{} &  \colhead{} 
}
  \tablehead{
\colhead{Model} & \colhead{} & \colhead{} & \colhead{Parameters} &  \colhead{} &  \colhead{} &  \colhead{} &  \colhead{} 
}
  \startdata
{\sc highecutpl}\tablenotemark{a} &$N_{\rm H}$  &$\Gamma$  &$E_{\rm c}$ (keV) &$E_{\rm f}$ (keV) &$\chi^{2}_{\nu}$ (dof) &$L_{\rm 0.5-10}$\tablenotemark{c} &$L_{\rm 0.5-100}$\tablenotemark{c} \\
		& 1.1$\pm{0.2}$ &    0.75$\pm{0.11}$ &   18$\pm{2}$  &  4$\pm{2}$  &  0.919 (157)  &  1.9 & 5.3 \\
\noalign{\smallskip\hrule\smallskip}
{\sc cutoffpl}\tablenotemark{a}   & $N_{\rm H}$  & $\Gamma$  & $E_{\rm c}$ (keV) &  & $\chi^{2}_{\nu}$ (dof) & &   \\
		& 0.76 $^{+0.18} _{-0.16}$    &  0.05$\pm{0.18}$ &  7.2$^{+1.2} _{-1.0}$  & & 0.989 (158) &  1.8 & 4.2 \\
\noalign{\smallskip\hrule\smallskip}
{\sc compTT}\tablenotemark{b} & $N_{\rm H}$   & $T_{\rm 0}$ (keV)      &$T_{\rm e}$ (keV)    & $\tau$      & $\chi^{2}_{\nu}$ (dof) & &  \\
		& 0.43 $^{+0.19} _{-0.15}$    &  0.80$\pm{0.14}$ &  4.3 $^{+0.5} _{-0.4}$    &  19$\pm{3}$  & 0.934 (157)  &  1.8  & 4.4 
\smallskip 
  \enddata 
  \tablenotetext{a}{$N_{\rm H}$ is the neutral hydrogen column density ($\times 10^{22}$ cm$^{-2}$), 
                          $\Gamma$ the power law photon index, $E_{\rm c}$ the cutoff energy (keV), $E_{\rm f}$
                         the exponential folding energy (keV).}
  \tablenotetext{b}{$T_{\rm 0}$ is the temperature of the Comptonized seed photons, $T_{\rm e}$ the temperature of the 
			Comptonizing electron plasma, $\tau$ the optical depth of the Comptonizing plasma (spherical geometry).}
  \tablenotetext{c}{In units of 10$^{36}$ erg s$^{-1}$ derived  assuming a distance of 3.6~kpc. 
			}
  \end{deluxetable}

\begin{deluxetable}{lrrrrrrr}
  \tabletypesize{\scriptsize}
  \tablewidth{0pc} 	      	
  \tablecaption{Spectral fits of simultaneous XRT and BAT data of \srcxte.\label{tab:specxte}}
  \tablehead{
\colhead{\srcigr} & \colhead{} & \colhead{} & \colhead{} &  \colhead{} &  \colhead{} &  \colhead{} &  \colhead{} 
}
  \tablehead{
\colhead{Model} & \colhead{} & \colhead{} & \colhead{Parameters} &  \colhead{} &  \colhead{} &  \colhead{} &  \colhead{} 
}
  \startdata
{\sc highecutpl}\tablenotemark{a} &$N_{\rm H}$  &$\Gamma$  &$E_{\rm c}$ (keV) &$E_{\rm f}$ (keV) &$\chi^{2}_{\nu}$ (dof) &$L_{\rm 0.5-10}$\tablenotemark{c} &$L_{\rm 0.5-100}$\tablenotemark{c} \\
		& 12.5 $^{+1.5} _{-4.3}$ &    1.4 $^{+0.5} _{-1.0}$ &   6 $^{+7} _{-6}$  &  16  $^{+12} _{-8}$ &  1.37 (53)  &  1.9  & 3.0 \\
\noalign{\smallskip\hrule\smallskip}
{\sc cutoffpl}\tablenotemark{a}   & $N_{\rm H}$  & $\Gamma$  & $E_{\rm c}$ (keV) &  & $\chi^{2}_{\nu}$ (dof) & &   \\
		& 11.9 $^{+3.9} _{-2.8}$    &  1.0$\pm{0.7}$    &  13$^{+14} _{-5}$  & & 1.36 (54) &  1.6 & 3.1 \\
\noalign{\smallskip\hrule\smallskip}
{\sc compTT}\tablenotemark{b} & $N_{\rm H}$   & $T_{\rm 0}$ (keV)      &$T_{\rm e}$ (keV)    & $\tau$      & $\chi^{2}_{\nu}$ (dof) & &  \\
		& 8.2  $^{+5.9} _{-2.4}$   &  1.3 $^{+0.4} _{-1.3}$ &  8 $^{+16} _{-3}$    &  6.8$^{+2.5} _{-6.1}$  & 1.37 (53)  &  1.1  & 2.2
\smallskip 
  \enddata 
  \tablenotetext{a}{$N_{\rm H}$ is the neutral hydrogen column density ($\times 10^{22}$ cm$^{-2}$), 
                          $\Gamma$ the power law photon index, $E_{\rm c}$ the cutoff energy (keV), $E_{\rm f}$
                         the exponential folding energy (keV).}
  \tablenotetext{b}{$T_{\rm 0}$ is the temperature of the Comptonized seed photons, $T_{\rm e}$ the temperature of the 
			Comptonizing electron plasma, $\tau$ the optical depth of the Comptonizing plasma (spherical geometry).}
  \tablenotetext{c}{In units of 10$^{36}$ erg s$^{-1}$ derived  assuming a distance of 2.7~kpc. 
						}
  \end{deluxetable}

\end{document}